\documentclass[11pt]{article}
\usepackage{cite}
\usepackage{amsmath,amsfonts,amssymb}
\usepackage[small,bf,hang]{caption}
\usepackage{slashed}
\usepackage{color}

\usepackage{extarrows}
\usepackage{hyperref}

\usepackage{geometry}
\usepackage{parskip} 

\def\hybrid{
        \topmargin -20pt
        \oddsidemargin 0pt
        \headheight 0pt \headsep 0pt
        \textwidth 6.25in 
        \textheight 9.5in 
        \marginparwidth .875in
        \parskip 5pt plus 1pt \jot = 1.5ex}

\hybrid

 \csname
@addtoreset\endcsname{equation}{section}

\newcommand{\Sgen}{\mathcal{S}}

\def\moth{\mathsurround=0pt}
\newdimen\zo \zo=0pt

\def\tick{\leaders\hrule height 0.5ex depth 0pt \hskip 0.5pt}
\def\upboxfill{$\moth \setbox\zo\hbox{\tick}%
  \hskip 3pt\hbox to 0pt{$\tick$\hss}\hrulefill \hbox to 7.5pt{$\tick$\hss}$}

\def\dtick{\leaders\hrule height .34pt depth 0.5ex \hskip 0.5pt}
\def\downboxfill{$\moth \setbox\zo\hbox{\dtick}%
  \hskip 2pt\hbox to 0pt{$\dtick$\hss}\hrulefill \hbox to 2pt{$\dtick$\hss}$}

\def\bec{\begin{center}}
\def\ec{\end{center}}

\def\L{\Lambda}

\def\S{{\cal S}}

\def\cdS{{\dot{\cal S}}}

\def\nn{\nonumber}

\def\be{\begin{equation}}
\def\ee{\end{equation}}
\def\bea{\begin{eqnarray}}
\def\eea{\end{eqnarray}}
\def\ba{\begin{array}}
\def\ea{\end{array}}

\newcommand{\tr}[1]{\text{Tr}(#1)}
\newcommand{\Tr}[1]{\text{Tr}\left(#1\right)}

\renewcommand{\Sgen}{\mathcal{S}}

\newcommand{\E}{\mathbb{E}}

\renewcommand{\L}{\mathbb{L}}

\begin{document}

\begin{titlepage}
\rightline{}
\rightline{June 2021}
\rightline{HU-EP-21/16-RTG} 
\begin{center}
\vskip 1.5cm
 {\Large \bf{   
 General String Cosmologies at Order $\alpha'{}^{\,3}$}}
\vskip 1.7cm

{\large\bf {Tomas Codina$^\dag$, Olaf Hohm$^\dag$ and Diego Marques$^*$}}
\vskip 1cm

$^\dag$ {\it   Institute for Physics, Humboldt University Berlin,\\
 Zum Gro\ss en Windkanal 6, D-12489 Berlin, Germany}\\
 
\vskip .3cm

$^*$ {\it   Instituto de Astronom\'ia y F\'isica del Espacio, \\
 Casilla de Correo 67 - Suc. 28 (C1428ZAA), Buenos Aires, Argentina}\\
\vskip .1cm

\vskip .4cm

tomas.codina@physik.hu-berlin.de, ohohm@physik.hu-berlin.de, diegomarques@iafe.uba.ar

\vskip .4cm

\end{center}

\bigskip\bigskip
\begin{center} 
\textbf{Abstract}

\end{center} 
\begin{quote}

We compute  the cosmological  reduction of 
general string theories, including bosonic, heterotic and type II string theory  to  order $\alpha'^{3}$, i.e., with up to eight derivatives. 
To this end we refine recently introduced methods that allow one to bring the reduced theory in one dimension to a canonical form 
with only first-order time derivatives. The resulting theories are compatible with a continuous $O(d,d,\mathbb{R})$ invariance, 
which in turn fixes the B-field couplings.

\end{quote} 
\vfill
\setcounter{footnote}{0}
\end{titlepage}

\section{Introduction}

One of the fascinating features  of string theory is its  invariance under  dualities  that, in the
simplest case, send the metric $g$ to its inverse $g^{-1}$. For string backgrounds with $d$-dimensional translation invariance
such dualities belong to the group  $O(d,d,\mathbb{R})$ \cite{Meissner:1991zj,Meissner:1991ge,Sen:1991zi}.
In particular, the theory for cosmological string backgrounds, with fields depending only on time,
is invariant under $O(9,9)$  in the case of superstring theory and under $O(25,25)$   in the case of bosonic string theory.
For Friedmann-Lemaitre-Robertson-Walker backgrounds this includes the transformation that sends the scale
factor of the universe to its inverse, a fact that immediately inspires  ideas of how to apply  string theory in cosmology, see e.g.~\cite{Brandenberger:1988aj,Tseytlin:1991xk,Veneziano:1991ek}. It is challenging, however,  to upgrade such ideas to fully reliable
string cosmology proposals. One reason is that even classical string theory restricted to the massless fields contains an infinite number of higher-derivative
$\alpha'$ corrections which contribute to the cosmological equations, and only very little is known about these corrections. (See \cite{Green:2011cn} for applications of higher derivative corrections
in string cosmology.)
In this paper, which is a continuation of our recent letter \cite{Codina:2020kvj},
we determine the cosmological reduction for all string theories up to and including $\alpha'^3$ (i.e.~with up to eight derivatives) for metric, B-field
and dilaton.

Our analysis is made possible by the results of \cite{Hohm:2019jgu,Hohm:2015doa}, which classify the  $\alpha'$ corrections in one dimension (cosmic time)
up to field redefinitions. This leads to a drastic reduction of the number of possible terms arising in the one-dimensional action.
It has been known since the seminal work in \cite{Tseytlin:1991wr,Meissner:1996sa} that the $O(d,d)$ transformations themselves
receive $\alpha'$ corrections when written in terms of standard supergravity field variables, but in one dimension these corrections can be removed by suitable field redefinitions,
so that one can test directly for $O(d,d)$ invariance by passing to a canonical field basis.\footnote{For dimensional reduction to a generic number of dimensions, however, the $O(d,d)$ transformations receive non-trivial Green-Schwarz-type  $\alpha'$-deformations  \cite{Eloy:2019hnl}, which has a precursor in double field theory 
\cite{HSZ,Hohm:2014xsa,Marques:2015vua,Baron:2017dvb}, see also earlier work in \cite{Bergshoeff:1995cg}.}
The theory can then be written in terms of conventional $O(d,d)$
covariant fields, notably the generalized metric
\begin{equation}
	\Sgen \equiv  \begin{pmatrix}
	b g^{-1} & g - b g^{-1} b\\
	g^{-1} & - g^{-1} b
	\end{pmatrix} \ ,
	\end{equation}
that takes values in $O(d,d)$. Here $g$ and $b$ denote the spatial components of the redefined metric and B-field.
The results of \cite{Hohm:2019jgu} imply that the cosmological action for any string theory can be brought
to a manifestly $O(d,d)$ covariant form that to order ${\cal O}(\alpha'{}^3)$ reads
\be\label{MinimalActionCoeff}
 \begin{split}
S  = \int dt \, e^{- \Phi} \Big[&- \dot \Phi^2 - \frac 1 8  \tr{\cdS^2} \\
&+ \alpha' c_{2,0} \tr{\cdS^4} + \alpha'{}^2 c_{3,0} \tr{\cdS^6} \\
&+ \alpha'{}^3 \left(c_{4,0} \tr{\cdS^8} + c_{4,1} \tr{\cdS^4}^2\right) \vphantom{\frac 1 8}\Big] \ .
 \end{split}
 \ee
In this paper we develop a systematic procedure to bring any action dimensionally reduced to one dimension to a form that contains only
first order time derivatives. We will ignore all B-field couplings in the dimensional reduction, which is sufficient in order to determine the above coefficients,
but then  $O(d,d)$ guarantees that the B-field couplings will be as implied by the general action (\ref{MinimalActionCoeff}).

It is indeed guaranteed on general grounds that $O(d,d)$ is preserved to all orders in $\alpha'$ \cite{Sen:1991zi}, and this will be confirmed here.
Even after truncating the B-field a $\mathbb{Z}_2$ symmetry is left, which in turn poses strong constraints on the purely gravitational couplings that can exist
in higher  dimensions. In \cite{Codina:2020kvj} we showed that the $\alpha'^3$ corrections involving the eight-derivative couplings known as $t_8t_8R^4$ and
$\epsilon_{10}\epsilon_{10}R^4$ need to arise with a specific relative coefficient in order to be compatible with $O(d,d)$, which turns out
to be the coefficient previously determined by other methods \cite{GrisaruZanon}. Similarly, we find here that the Riemann-cube terms known to arise in bosonic
string theory at order $\alpha'^2$ need to be accompanied by a Gauss-Bonnet-type combination at the same order, with a relative coefficient that again
confirms previous results in \cite{Metsaev:1986yb}. Apart from type II string theory, whose corrections start only at order $\alpha'^3$ and have been analyzed in \cite{Codina:2020kvj},
we will analyze here bosonic and heterotic string theory but also the Hohm-Siegel-Zwiebach (HSZ) theory constructed in \cite{HSZ}.
We find for the four free coefficients in (\ref{MinimalActionCoeff}) that are not fixed by $O(d,d)$ the following values:
\medskip
\begin{center}
	\begin{tabular}{|| c || c | c | c | c ||}
		\hline
		& $c_{2,0}$ & $c_{3,0}$ & $c_{4,0}$ & $c_{4,1}$ \\ [0.5ex]
		\hline\hline
		Bosonic &  $\frac 1 {2^6}$  & $- \frac 1 {3 . 2^7}$ &  $\frac 1 {2^{12}}  - \frac 3 {2^{12}}  \zeta(3)$  &  $ \frac{1} {2^{16}}  + \frac 1 {2^{12}} \zeta(3) $\\
		\hline
		HSZ & $0$ & $\frac 1 {3 . 2^7}$  & $0$ & $0$ \\
		\hline
		Heterotic & $\frac 1 {2^7}$ & $0$  & $ - \frac 3 {2^{12}} \zeta(3)$ & $- \frac {15}{2^{19}}  + \frac 1 {2^{12}} \zeta(3)$\\
		\hline
		Type II & $0$ & $0$  & $ - \frac 3 {2^{12}} \zeta(3)$ & $\frac 1 {2^{12}} \zeta(3)$\\
		\hline
	\end{tabular}
\end{center}

\bigskip

We find it instructive to give this result in an  alternative form, in terms of three parameters $(a,b,c)$ that encode the $\alpha'$ corrections
of all string theories. Specifically, for the theories considered here these parameters encode $\alpha'$ via the following table:
\medskip
\begin{center}
	\begin{tabular}{|| c || c | c | c ||}
		\hline
		& $a$ & $b$ & $c$ \\ [0.5ex]
		\hline\hline
		Bosonic &  $\alpha'$  & $\alpha'$ & $\alpha'$ \\
		\hline
		HSZ & $-\alpha'$ & $\alpha'$ & $0$  \\
		\hline
		Heterotic & $0$ & $\alpha'$ & $\alpha'$ \\
		\hline
		Type II & $0$ & $0$ & $\alpha'$ \\
		\hline
	\end{tabular}
\end{center}
\medskip
The cosmological action of the form (\ref{MinimalActionCoeff}) that we find here can then be written as
\begin{eqnarray}
S \!&=&\! \int dt \, e^{- \Phi} \, \left[- \dot \Phi^2 - \frac 1 8  \tr{\cdS^2} + \frac {(a+b)} {2^7} \tr{\cdS^4} - \frac {a b} {3 \cdot 2^7} \tr{\cdS^6}
\right. \nn \\&& \ \ \ \ \ \ \ \ \ \ \ \ \ \ \   +\, \frac {ab (a+b)} {2^{13}} \tr{\cdS^8}   +  \left(\frac {1} {2^{13}} ab (a + b) - \frac {15}{2^{19}} (a + b)^3 \right) \tr{\cdS^4}^2  \label{BiParametricCosmo}
\\&& \ \ \ \ \ \ \ \ \ \ \ \ \ \ \ \left. +\, c^3 \frac {\zeta(3)} {2^{12}} \left( -3 \tr{\cdS^8}   + \tr{\cdS^4}^2\right) \vphantom{\frac 1 8}\right]\ .  \nn
\end{eqnarray}

This parametrization of the action is motivated by double field theory (a reformulation of the target space theory that is $O(d,d)$
covariant before dimensional reduction), which permits a 2-parameter $\alpha'$-deformation \cite{Hohm:2014xsa,Marques:2015vua} that  is invariant under the $\mathbb{Z}_2$
that  exchanges the two parameters $a, b$ and simultaneously sends the $O(d,d)$ metric $\eta$ to $-\eta$. The above parametrization has been chosen to reflect the same
symmetry (otherwise we could add ${\cal O}(\alpha'{}^3)$ terms proportional to $a (a + b) (a - b)$, since this combination vanishes for all of the above theories).
As an aside we emphasize that since the classification of \cite{Hohm:2019jgu,Hohm:2015doa} yields only even
powers of $\dot{\cal S}$ the cosmological reduction of any string theory is $\mathbb{Z}_2$ invariant, and so $\mathbb{Z}_2$ odd contributions in higher dimensions
(as present in Green-Schwarz deformations) cannot survive cosmological reduction.
Finally, the third parameter $c$ appearing above is expected to indicate the presence of a new $\alpha'$-deformation
of double field theory to incorporate the fourth powers of the Riemann tensor with transcendental coefficient  $\zeta(3)$ (see
\cite{Hronek:2020xxi} for the challenges that arise when trying to define this new deformation).

The remainder of this paper is organized as follows. In Section 2 we explain in detail the systematic procedure that is used in order to bring
the dimensionally reduced actions into canonical form. This method will then be applied in  Section 3 to the various string theories.
In Section 4 we close with a brief outlook.

\vspace{.2cm}

\noindent \textit{Note added:} After the submission of the first version of this paper to the arxiv, ref.~\cite{Garousi:2021ocs} appeared, in which 
the coefficient $c_{3,0}$ for the bosonic string is computed including all B-field and dilaton couplings, in perfect agreement with the value for $c_{3,0}$ 
found here.

\section{General Approach}\label{sec_general}

In this section we introduce the general algorithmic procedure that brings the dimensionally reduced actions into a canonical  form, 
in which in particular only first-order time derivatives appear. This is a refinement of the methods introduced in \cite{Hohm:2015doa,Hohm:2019jgu}.  
In the subsequent sections this method will be applied to the various 
string theories. Concretely, for each string theory we start with the low-energy effective action in $D=10$ or $D=26$ dimensions including higher derivative corrections 
up to and including order $\alpha'{}^3$, written schematically as 
\begin{equation}\label{action10d}
S = \int d^{D} x \sqrt{-G} e^{-2 \phi}\left[ \mathcal{L}^{(0)} + \alpha' \mathcal{L}^{(1)} + \alpha'{}^2 \mathcal{L}^{(2)} + \alpha'{}^3 \mathcal{L}^{(3)} \right]\;. 
\end{equation}
We will simplify the analysis by setting the Kalb-Ramond $B$-field to zero, $B=0$. This is sufficient for our purposes --- precisely because 
the $O(d,d)$ duality, with $d = D - 1$, allows us to reconstruct the $B$-field couplings. 
Under this assumption the  leading term in the above action, which is common to all string theories, 
is given by
\begin{equation}
\mathcal{L}^{(0)} = R + 4 \nabla_\mu \phi \nabla^\mu \phi\;. 
\end{equation}

Let us then turn to the general procedure of cosmological reduction. 
We take the $D$-dimensional metric $G_{\mu\nu}$ or, equivalently, the $D$-dimensional  vielbein $e_{\mu}{}^{\alpha}$ and the dilaton $\phi$ to depend only on time, 
making  the ansatz
\begin{equation}\label{Cosmoansatz}
e_{\mu}{}^{\alpha} = \text{diag}(n,e_i{}^a) \;, \qquad G_{\mu \nu} = \text{diag}\left(-n^2, g_{i j}\right) \;, \qquad \phi = \frac{1}{2} \Phi + \frac{1}{2} \log\left(\sqrt{g}\right)\;, 
\end{equation}
in terms of spatial metric $g_{ij}$, the dilaton $\Phi$ and the lapse function $n$. 
This ansatz is used in the $D$-dimensional action, truncating all derivatives but the time derivative. 
The actions we consider in the following contain the Riemann tensor, possibly Chern-Simons terms for the Levi-Civita connection, and dilaton couplings. 
In contradistinction to any analysis at a fixed order in $\alpha'$, here in principle we have to keep track of couplings containing Ricci tensors, Ricci scalars 
and dilaton contributions. 
However, once we reach order $\alpha'^3$ the couplings containing Ricci tensors and Ricci scalars 
can be eliminated by field redefinitions, at the cost of introducing further dilaton couplings. 
In \cite{Codina:2020kvj} we showed, at order $\alpha'^3$,  that under cosmological reduction derivatives of the dilaton can either be removed 
by field redefinitions or else violate $O(d,d)$ duality invariance. Since here we assume $O(d,d)$ invariance we will neglect 
all dilaton derivatives at order $\alpha'^3$.

We use the following conventions for the Levi-Civita connection, the Riemann tensor and Chern-Simons terms: 
\begin{equation}\label{definitions}
\begin{aligned}
\omega_{\mu \alpha}{}^{\beta} &= e_\alpha{}^\nu \nabla_{\mu} e_\nu{}^\beta = e_\alpha{}^\nu \partial_{\mu} e_\nu{}^\beta - e_\alpha{}^\nu \Gamma_{\mu \nu}{}^{\rho} e_{\rho}{}^\beta
\,, \\
R^{\rho}{}_{\sigma \mu \nu} &= \partial_\mu \Gamma_{\nu \sigma}{}^\rho - \partial_\nu \Gamma_{\mu \sigma}{}^\rho + \Gamma_{\mu \lambda}{}^\rho \Gamma_{\nu \sigma}{}^\lambda - \Gamma_{\nu \lambda}{}^\rho \Gamma_{\mu \sigma}{}^\lambda\, ,\\
R_{\mu \nu \alpha}{}^{\beta}(\omega) &= \partial_\mu \omega_{\nu \alpha}{}^\beta - \partial_\nu \omega_{\mu \alpha}{}^\beta + \omega_{\mu \alpha}{}^\gamma \omega_{\nu \gamma}{}^\beta - \omega_{\nu \alpha}{}^\gamma \omega_{\mu \gamma}{}^\beta = - e_{\alpha}{}^\sigma e_\rho{}^{\beta} R^{\rho}{}_{\sigma \mu \nu}\, , \\
\Omega_{\mu \nu \rho}(\omega) &= \Tr{\omega_{[\mu} \partial_{\nu} \omega_{\rho]} + \frac{2}{3} \omega_{[\mu} \omega_{\nu} \omega_{\rho]} }\;, 
\end{aligned}
\end{equation} 
where $\Gamma_{\mu\nu}{}^{\rho}$ are the familiar Christoffel symbols. 
Inserting the cosmological reduction ansatz (\ref{Cosmoansatz}) in here one obtains 
for the non-vanishing components:  
\begin{equation}\label{splitcomponentss}
\begin{aligned}
\Gamma_{0 i}{}^{j} &= \frac{n}{2} L_i{}^j \;, \qquad\quad\quad    \Gamma_{i j}{}^{0} = \frac{1}{2 n} L_{i j} \;, \qquad\quad\quad    \Gamma_{0 0}{}^0 = \dot{n}\;, \\
\omega_{i a}{}^{\bar{0}} &= -\frac{1}{2} L_{i a} \;, \qquad\quad  \; \omega_{0 a}{}^{b} = n \, e_a{}^j \dot{e}_{j}{}^b -\frac{n}{2} L_{a}{}^b\;, \\
R_{i j k l} &= \frac{1}{2} L_{k[i} L_{j] l} \;, \qquad R_{0 i 0 j} = - \frac{n^2}{2}\dot{L}_{i j} - \frac{n^2}{4} L^2_{i j}\;, \\
\Omega_{0 i}{}^j(\omega) &= \frac{n}{6} \dot{L}_{[i}{}^k L_{j] k}\;, \qquad 
\nabla_0 \phi = \frac{n}{2} \dot{\Phi} + \frac{n}{4} (L)\;, 
\end{aligned}
\end{equation}
where we split the flat index as $\alpha=(\bar{0},a)$.  
Here the dots denote   time derivatives $\dot{\psi} \equiv  \frac{1}{n} \partial_t \psi$ that are reparametrization invariant, while parenthesis $(\;)$ denote traces of $d\times d$ matrices 
such as  $L \equiv \dot{g} g^{-1}$. Internal indices are raised and lowered with $g$, namely $L_{i j} \equiv L_i{}^k g_{k j}$, $\dot{L}_{i j} \equiv \dot{L}_i{}^k g_{k j}$ and flattened with $e_{i}{}^a$ such that $L_{i a} = L_{i}{}^j g_{j k} e_{a}{}^k$.

We now explain the step-by-step procedure of bringing the dimensionally reduced action to the canonical form. 
The first step, which may be technically tedious but is conceptually straightforward,  
consists of inserting (\ref{splitcomponentss}) into the $D$-dimensional action in order to obtain the cosmological reduction. 
This yields a one-dimensional theory with an action of the  form 
\begin{equation}\label{action1d}
S = \int d t \, n \, e^{-\Phi} \left[ L^{(0)} + \alpha' L^{(1)} + \alpha'{}^2 L^{(2)} + \alpha'{}^3 L^{(3)} \right]\;. 
\end{equation}
In here the leading term is common to all string theories and given by 
\begin{equation}
L^{(0)} = - \dot{\Phi}^2 + \frac{1}{4} (L^2)\;. 
\end{equation}
The other terms depend on the string theory under consideration. An important observation, which follows from our above assumptions and the form of (\ref{splitcomponentss}), 
is that all these terms are built from traces of different products of $L$ and its time derivatives. In the next step one exploits all possible field redefinitions.   
We need to compute the general variation of (\ref{action1d}) that defines the equations of motion:  
\begin{equation}\label{generalVARaction}
\delta S = \int d t \, n \, e^{- \Phi} \left[ \frac{1}{2} \Tr{\left(E_g + E_g^t\right) \delta g} + E_n \frac{\delta n}{n} + E_\Phi \delta \Phi\right] \;. 
\end{equation}
As for the action, the equations of motion have an $\alpha'$ expansion. Denoting the fields collectively by 
 \be
  \Psi \equiv \{ g, n, \Phi \}\;, 
 \ee
we write the $\alpha'$ expansion of the equations of motion as 
\begin{equation}\label{EOM}
E_\Psi = E^{(0)}_\Psi + \alpha' E^{(1)}_\Psi + \alpha'{}^2 E^{(2)}_\Psi + \mathcal{O}(\alpha'{}^3) = 0 \;. 
\end{equation}
The terms of order $\alpha'{}^3$ or higher will not be needed in this paper. 
Again, the lowest-order terms are the same for all string theories and are given by 
\begin{equation}\label{EOM0}
\begin{aligned}
E_g^{(0)} &= \frac{1}{2} \dot{\Phi} L - \frac{1}{2} \dot{L}\;, \\
E_n^{(0)} &= \dot{\Phi}^2 - \frac{1}{4} (L^2)\;, \\
E_\Phi^{(0)} &= 2 \ddot{\Phi} - \dot{\Phi}^2 - \frac{1}{4} (L^2)\;, 
\end{aligned}
\end{equation}
while the higher order contributions depend on the string theory under consideration. 
As long as we recall the equations of motion for $n$ (and the freedom to perform field redefinitions of $n$) we can 
gauge fix reparametrization invariance by setting $n=1$. In this case the dot reduces to the ordinary time derivative, $\dot{\Psi} = \partial_t \Psi$. 
We may always restore time reparametrization invariance simply by reinterpreting the dot as $\frac{1}{n}\partial_t$.

Let us now consider a general field redefinition 
\be
 \Psi \rightarrow  \Psi' = \Psi + \delta \Psi \;, 
\ee
which we take to be perturbative in $\alpha'$, so that we can expand 
 \be
  \delta \Psi = \alpha' \delta^{(1)} \Psi + \alpha'{}^2 \delta^{(2)} \Psi + \alpha'{}^3 \delta^{(3)} \Psi + \cdots\;. 
 \ee
The action expands as follows: 
 \be
  S'[\Psi']  \equiv S[\Psi+\delta \Psi] = S[\Psi] + \Delta_1S \cdot \delta \Psi 
  +\frac{1}{2} \Delta_2S \cdot (\delta \Psi)^2 + \frac{1}{3!} \Delta_3S \cdot  (\delta \Psi)^3+\cdots\;, 
 \ee 
 where we use a symbolic notation in which the integral is not displayed explicitly.  
 This equation defines implicitly the $n$-th variational derivatives $\Delta_n S\equiv \frac{\delta^nS}{\delta \Psi^n}$, 
 the first of which, in agreement with our notation above, is also written as 
  \be
   \Delta_1S\equiv \frac{\delta S}{\delta \Psi} \equiv E_{\Psi} \,,
  \ee
with $\alpha'$ expansion (\ref{EOM}).   
Similarly, we write the $\alpha'$ expansion of $\Delta_nS$ as 
 \be
  \Delta_nS = \Delta_nS^{(0)} + \alpha' \Delta_nS^{(1)} + (\alpha')^2 \Delta_nS^{(2)}+\cdots \;. 
 \ee
 The redefined action $S'$  
 can then be written as 
\be\label{fieldredef}
\begin{split}
S' = S^{(0)}
&+ \alpha' \Big(S^{(1)} + E_\Psi^{(0)} \cdot \delta^{(1)} \Psi\Big)\\
&+ \alpha'{}^2\Big(S^{(2)} + E_\Psi^{(1)} \cdot \delta^{(1)} \Psi  + E_\Psi^{(0)}  \cdot \delta^{(2)} \Psi 
+\underline{\underline{\frac{1}{2} \Delta_2S^{(0)} \cdot \big(\delta^{(1)}\Psi\big)^2}} \Big)\\
&+ \alpha'{}^3\Big(S^{(3)} + E_\Psi^{(2)} \cdot  \delta^{(1)} \Psi  + E_\Psi^{(1)} \cdot \delta^{(2)} \Psi + E_\Psi^{(0)} \cdot \delta^{(3)} \Psi \\
& 
+\underline{\frac{1}{2} \Delta_2S^{(1)} \cdot \big(\delta^{(1)}\Psi\big)^2 
+\Delta_2S^{(0)} \cdot \delta^{(1)}\Psi\cdot  \delta^{(2)}\Psi + \frac{1}{3!} \Delta_3S^{(0)} \cdot \big(\delta^{(1)}\Psi\big)^3} \Big) + \cdots\;. 
\end{split}
\ee 
The natural method of bringing the action into a canonical form then proceeds order-by-order in $\alpha'$: one first picks a $\delta^{(1)} \Psi$ to bring the 
action to first order in $\alpha'$ to canonical form, but this in turn induces new terms proportional to $E_\Psi^{(1)}$ 
and $\Delta_2S^{(0)}$ into the action of second order in $\alpha'$. 
These can then be brought to a canonical form by picking a suitable $\delta^{(2)} \Psi$. Both  $\delta^{(1)} \Psi$ and $\delta^{(2)} \Psi$ then induce new terms 
into the action of third order in $\alpha'$, which finally can be brought to a canonical form by picking a suitable $\delta^{(3)} \Psi$. 

In principle, the above procedure requires that one keeps track of the field redefinitions  $\delta^{(1)} \Psi$, $\delta^{(2)} \Psi$ and $\delta^{(3)} \Psi$, 
which can become rather tedious. We will now show, however, that for a large class of contributions to the redefined action there is a simplified 
procedure for which one does not need to keep track of the field redefinitions. These correspond to  the terms in (\ref{fieldredef}) that are not underlined. 
For the underlined terms, on the other hand,  it is necessary to keep track of the field redefinitions, which depend on the theory, but we will see below  that 
only the explicit form of $\delta^{(1)} \Psi$ is needed whose  
contribution takes a universal form for all 
string theories.

In order to explain the procedure let us first consider only the terms in the redefined action that are not underlined 
and explain how they can be brought to a canonical form upon choosing appropriate $\delta\Psi$. 
In this case field redefinitions amount  to the use of 
equations of motion in the action, including contributions to higher order in $\alpha'$.
In order to make this concrete let us suppose that the action to first order in $\alpha'$ contains a term multiplying the lowest order equations of motion, i.e., 
 \be\label{S1example}
  S^{(1)} =  X(\Psi)\cdot E_{\Psi}^{(0)} + \cdots \;, 
 \ee
where $X(\Psi)$ is an arbitrary function of the fields $\Psi$ (of second order in derivatives) and the ellipsis denote the remaining terms in the action. 
Consider now a field redefinition with 
 \be
  \delta^{(1)}\Psi = -X(\Psi)\;. 
 \ee
From (\ref{fieldredef}) we then infer that in the redefined action $S'$ the term in (\ref{S1example}) is eliminated. More precisely, 
 \be
 \begin{split}
  S' = &\,-\alpha' X(\Psi)\cdot \big(\alpha' E_{\Psi}^{(1)}+\alpha'^{2} E_{\Psi}^{(2)}\big)+\cdots
  \;, 
 \end{split}
 \ee
where the ellipsis denote the same terms as in  (\ref{S1example}), which are unaffected by the redefinition, 
(and we recall that we neglected the underlined terms in (\ref{fieldredef}), to which we return soon). 

 The upshot is that in the action (\ref{S1example}) we may simply use the equations of motion (\ref{EOM}) in the form 
 $E_{\Psi}^{(0)} = -\alpha' E_{\Psi}^{(1)}-\alpha'^2 E_{\Psi}^{(2)}+\cdots$, where the higher order terms can be ignored as we are only interested 
 in contributions up to and including $\alpha'^3$. 
 One can then proceed order-by-order in $\alpha'$ by freely using the equations of motion in the action at each order in $\alpha'$, 
 as long as one keeps track 
 of the  terms induced to next order in $\alpha'$. 
To this end we will use the equations of motion (EOM) in the form (c.f.~(\ref{EOM0})) 
\begin{subequations}
\label{EOMsimplifieds}
\begin{align}
\dot{L} &= \dot{\Phi} L + \alpha' \Delta_g\;,  \label{EOML}\\
\dot{\Phi}^2 &= \frac{1}{4} (L^2) + \alpha' \Delta_n \;,\label{EOMn}\\
\ddot{\Phi} &= \frac{1}{2} \dot{\Phi}^2 + \frac{1}{8} (L^2) + \alpha' \Delta_\Phi \;, \label{EOMPhi}
\end{align}
\end{subequations}
where $\alpha'\Delta_\Psi$ denotes a generic correction starting at order $\alpha'$. Again, its  particular form depends on the string theory considered 
and will be computed in the subsequent sections.

Next we give a step-by-step procedure  to use field redefinitions in order  to remove, at any given order in $\alpha'$, 
any appearance of $\dot{L}, \dot{\Phi}$ and $(L^2)$ and their time derivatives, at the expense of inducing new terms at higher order in $\alpha'$. 
In the following we will refer to terms containing $\dot{L}, \dot{\Phi}$ and $(L^2)$ and their time derivatives  as removable.  
\begin{enumerate}
	\item Beginning with the first order action $\alpha'S^{(1)}$ the first step is to use repeatedly (\ref{EOML}) and its derivatives  to eliminate $\dot{L}$ and its derivatives. 
	For instance, if the action contains no higher derivatives than $\dot{L}$ 
	this yields a new action at order $\alpha'$, 
	with $\dot{\Phi} L$ substituted for $\dot{L}$, and induced terms for the next two orders involving  $\alpha'\Delta_g$: 
	\begin{equation}
	\alpha'S^{(1)}(\dot{L}) = \alpha'S^{(1)}(\dot{\Phi} L + \alpha' \Delta_g) = \alpha'S^{(1)}(\dot{\Phi} L) + \mathcal{O}(\alpha'{}^2) \ .
	\end{equation}
	At this point, everything in $S^{(1)}$ is written in terms of traces of products of $L$ and powers and derivatives of $\dot{\Phi}$.
	If, on the other hand, the action contains higher derivatives than $\dot{L}$ the above procedure has to be repeated until the 
	action depends only on traces of products of $L$ (and dilaton terms). 
	
	\item Next, we eliminate any higher power of $\dot{\Phi}$ by using (\ref{EOMn}) repeatedly. The result is the substitution of $\frac{1}{4}(L^2)$ for $\dot{\Phi}^2$, 
	at the current order, and some induced terms which will appear in the next orders: 
	\begin{equation}
	\alpha'S^{(1)}(\dot{\Phi}^2) = \alpha'S^{(1)}\big(\tfrac{1}{4}(L^2)\big) + \mathcal{O}(\alpha'{}^2)  \ .
	\end{equation}
	By this method, any even powers of $\dot{\Phi}$ can be removed, so now $S^{(1)}$ depends linearly on $\dot{\Phi}$, its higher derivatives and traces of products of $L$.
	
	\item Next, we eliminate any higher derivative of $\dot{\Phi}$ by using (\ref{EOMPhi}) repeatedly, in each step making the replacement 
		\begin{equation}
	\alpha'S^{(1)}(\ddot{\Phi}) = \alpha'S^{(1)}\big(\tfrac{1}{2}\dot{\Phi}^2 + \tfrac{1}{8}(L^2)\big) + \mathcal{O}(\alpha'{}^2) \ ,
	\end{equation}
	until we are left with only first-order time derivatives of the dilaton. 
	At the end of these iterations, this generates powers of $\dot{\Phi}$ and possibly  higher derivatives of ${L}$. One can eliminate higher derivatives of $L$ and 
	even powers of $\dot{\Phi}$ by repeating steps \textbf{1} and \textbf{2}. Since step \textbf{1} can also produce new $\ddot{\Phi}$ terms, this procedure might need to be iterated more than once. 
	Since in each step the number of derivatives of the dilaton decreases this procedure is guaranteed to terminate. 
	At the end $S^{(1)}$ contains at most a single $\dot{\Phi}$, together with products of traces of $L$, i.e.
	\begin{equation}
	\alpha' S^{(1)} = \alpha' \int dt \, e^{-\Phi} \left[\dot{\Phi} X(L) + Y(L)\right] + \mathcal{O}(\alpha'{}^2) \ ,
	\end{equation}
	where $X(L)$ and $Y(L)$ depend on traces of powers of $L$. In this case, the induced terms will generally depend on $\Delta_n$, $\Delta_\Phi$ and  $\Delta_g$.
	\item Now we can eliminate the dilaton contribution. We integrate by parts, 
	\be
	\alpha' \int dt \, e^{-\Phi} \dot{\Phi} X(L) = \alpha' \int dt \, e^{-\Phi} \dot{X}(L) \ ,
	\ee
	 and use (\ref{EOML}). At the current order, the $\dot{\Phi}L $ contribution gives the first term back but with a different coefficient. 
	 This yields an identity of the form 
	  \be
	   \alpha' \int dt \, e^{-\Phi} \dot{\Phi} X(L) = \alpha'{}^2 \int dt \, e^{-\Phi} F(\Delta_g)\;. 
	  \ee 
         At  this point $S^{(1)}$ does not contain any dilaton contribution so it is built just from traces of powers of $L$.
	
	\item We replace any appearance of $(L^2)$ in the action of the generic form $\alpha'\int dt \, e^{-\Phi} (L^2)X(L)$ 
	by the following procedure: Using (\ref{EOMn}) we replace $(L^2) \rightarrow 4 \dot{\Phi}^2 - 4 \alpha' \Delta_n $ and we integrate by parts one $\dot{\Phi}$ factor using $e^{-\Phi} \dot{\Phi}=- \partial_t \left(e^{-\Phi}\right)$. This creates terms with $\dot{L}$ and $\ddot{\Phi}$	which can be traded for $\dot{\Phi}^2$ and $(L^2)$ and higher orders by using (\ref{EOML}) and (\ref{EOMPhi}). Then, one replaces $\dot{\Phi}^2$ using (\ref{EOMn}). The final result produces the original integral but with a different coefficient, together with higher order corrections induced by $\Delta_g, \Delta_n$ and $\Delta_\Phi$. We thus obtain a relation of the form 
	\begin{equation}\label{actionSSTTEPP}
	\alpha'\int dt \, e^{-\Phi} (L^2)X(L) = \alpha'{}^2 \int dt \, e^{-\Phi} G(\Delta_g, \Delta_n, \Delta_\Phi)\;. 
	\end{equation}
	At this point, the first order action $S^{(1)}$ is written in the minimal basis proposed in \cite{Hohm:2019jgu}. 	

\end{enumerate}

After completing this procedure to first order in $\alpha'$ we can apply the algorithm to second order in $\alpha'$, taking as the starting point the dimensionally 
reduced action but supplemented by the terms that were induced by the field redefinitions of the previous iteration of the algorithm.  
This will yield an action including only traces of powers of $L$ (excluding $(L^2)$).   
Finally, we apply the algorithm to the action to third order in $\alpha'$, including  the induced terms to this order. 
In this final step we do not have to keep track of the induced $\Delta_{\Psi}$ terms such as in (\ref{actionSSTTEPP}), as these contribute only to fourth order 
in $\alpha'$.

The previous analysis showed how to bring the terms in the redefined action (\ref{fieldredef}) that are not underlined to a 
canonical form, which did not need the explicit form of $\delta\Psi$. 
We must now discuss 
how to bring the underlined terms to a canonical form, 
which will be seen to depend on the $\delta \Psi$. 
Importantly, however, we will show that one only needs to determine $\delta^{(1)} \Psi$ explicitly and that this takes a universal form. 
We first observe that  
the 
non-linear variations only  emerge from $S^{(0)}$ and $S^{(1)}$, which take a universal form for all string theories. 
Indeed, $S^{(0)}$ is the same for all string theories, while  the first order Lagrangian for the metric alone is given by 
\be
\mathcal{L}^{(1)} =  \frac {\gamma} 4 R_{\mu \nu \rho \sigma} R^{\mu \nu \rho \sigma} \ ,
\ee
with 
\medskip
\begin{center}
	\begin{tabular}{|| c | c | c | c | c ||}
		\hline
		  & Bosonic & HSZ & Heterotic & Type II  \\ [0.5ex]
		\hline
	$\gamma$ &  $1$  & $0$ & $\frac 1 2$ & $0$\\
		\hline
	\end{tabular}
\end{center}
\medskip
This fact  permits a unified treatment of the underlined terms. The direct cosmological reduction of the Lagrangian to this order is given by
\be
L^{(0)} + \alpha' L^{(1)}= - \dot \Phi^2 + \frac 1 4 (L^2) + \gamma \frac {\alpha'} 4 \left( \frac{1}{8} (L^4) + \frac{1}{8} (L^2)^2 +  (L^2 \dot{L}) + (\dot{L}^2)\right) \ , \label{FirstOrderCosmo}
\ee
and  the first order redefinitions required to remove the last three terms are given by
\begin{subequations}
\label{FirstOrderRedefs}
\begin{align}
\delta^{(1)} n &= \frac \gamma {32} (L^2) \;,\\
\delta^{(1)} \Phi &= \frac {3\gamma} {32} (L^2) \:, \\
G^{(1)} \equiv\delta^{(1)} g \,  g^{-1}  &= \gamma\, \left[ - \frac 1 4 \dot \Phi L + \frac 1 2 \dot L - \frac 1 4 L^2\right] \:.
\end{align}
\end{subequations}
Importantly, these are \textit{all} contributions to $\delta^{(1)}\Psi$.

Let us now see what the contributions of the underlined terms in (\ref{fieldredef}) are. We start with the terms with a single underline, which are of order $\alpha'{}^3$. 
The important observation is that all these terms can be verified to 
be entirely removable (i.e.~they contain $\dot{\Phi}$, $\dot{L}$ and $(L^2)$ or their time derivatives), and this is true for generic 
$\delta^{(2)}\Psi$. Thus, these terms 
do not contribute to the final canonical action to order $\alpha'^3$, since by removing them by field redefinitions (upon choosing an appropriate 
$\delta^{(3)}\Psi$) one only induces terms of higher order than $\alpha'^3$.

The only contribution that can affect the coefficients in the final canonical action 
is then 
the term $\Delta_2S^{(0)} \cdot \big(\delta^{(1)}\Psi\big)^2$ with a double underline in (\ref{fieldredef}), which takes the form
\begin{equation}\label{threeTerms}
\frac{1}{2} \Delta_2S^{(0)} \cdot \big(\delta^{(1)}\Psi\big)^2 = - \frac{3}{4}\,
(\dot G^{(1)}{}^2)\, -
(G^{(1)} \ddot G^{(1)})\,  + (G^{(1)} \dot G^{(1)})\,  \dot \Phi \, + \cdots\;. 
\end{equation}
In here all terms on the right-hand side can be verified to be removable to order $\alpha'^2$, where 
the dots denote terms whose removal leads to terms at order $\alpha'^3$ that 
are also removable at that order, so that they can be neglected completely. 
In contrast, the removal of the three terms explicitly written in (\ref{threeTerms})
induces non-trivial contributions to order $\alpha'^3$. 
It may be verified  that  one then induces 
a universal  non-removable third order contribution
\be
\Delta L^{(3)} = \frac {\gamma^3} {2^{11}} (L^4)^2  \ . \label{L42shift}
\ee
As a result, the only effect of the underlined terms to order $\alpha'{}^3$ is to shift the effective action by this quantity. One can then 
follow the procedure outlined above based on using $\alpha'$-corrected equations of motion, and add this single term at the end. Note that this will have no effect for Type II nor HSZ.

Once the field redefinition procedure has been completed the action to order $\alpha'{}^3$ is written in terms of the minimal basis containing the structures: $(L^4),(L^6), 
(L^8),(L^4)^2$ plus terms containing traces of odd powers of $L$.  The latter terms must, however, be absent as a consequence of duality invariance, c.f.~below.

The final step is to write  the theory, if possible, in terms of $O(d,d)$ covariant objects, using the following relation between $L_i{}^j$ and the generalized metric $\Sgen_M{}^N$
\begin{equation}
\begin{aligned}
\Sgen = \begin{pmatrix}
0 & g\\
g^{-1} & 0
\end{pmatrix}\;,  \qquad 
\dot \Sgen^{2 m} = \begin{pmatrix}
(-1)^m L^{2m} & 0\\
0 &  (-1)^m \left[L^{2m}\right]^t
\end{pmatrix} \;, 
\end{aligned}
\end{equation}
which implies 
\begin{equation}\label{SandL}
\tr{\dot \Sgen^{2 m}} = (-1)^m \,2 \, \Tr{L^{2m}}\;. 
\end{equation}
We see that $O(d,d)$ invariance requires that   odd powers  such as $(L^3)^2$ and $(L^3)(L^5)$ are actually absent, which in turn poses constraints on 
the $D$-dimensional higher-derivative corrections.

\section{Cosmological Reduction}

\subsection{Type II strings}

In this section we revisit the cosmological reduction of Type II strings up to order $\alpha'{}^3$  \cite{Codina:2020kvj} (see \cite{Moura:2007ks} for the reduction of these corrections to four dimensions). The action contains no order $\alpha'$ nor $\alpha'{}^2$ deformations. The corrections to order $\alpha'{}^3$  were computed from four-point scattering amplitudes in \cite{GrossWitten}, and later from the sigma-model beta function in \cite{GrisaruVenZanon, GrisaruZanon, FreemanPopeSohniusStelle}. Given the general form of the action (\ref{action10d}), we have\footnote{Here we rescaled $\alpha'$ in order to match conventions with the other strings. More precisely, we send $\alpha'$ in \cite{Codina:2020kvj} to $2 \alpha'$ here. After some change of conventions for $t_8$ and $\alpha'$, this result is in agreement with \cite{Gross:1986mw}. Without any change of conventions, (\ref{LS}) is in agreement with \cite{GrisaruZanon} up to a sign. It was explained in \cite{Jack:1989vp} that this sign should be a misprint (see discussion around equation (5.10) in \cite{Jack:1989vp}), so the correct value should be the one  in (\ref{LS}).}.
\begin{equation}\label{LS}
\begin{aligned}
\mathcal{L}_{\rm II}^{(1)} &= \mathcal{L}_{\rm II}^{(2)} = 0 \ ,\\
\mathcal{L}_{\rm II}^{(3)} &= \frac {\zeta(3)}{3.2^{11}} J(1) =  - \frac{\zeta(3)}{32} \left[ R^{\alpha \beta \mu \nu} R_{\mu \nu}{}^{\gamma \delta}  R_{\alpha \gamma}{}^{\rho \sigma} R_{\rho \sigma \beta \delta} - 4 R_{\alpha \beta}{}^{\gamma \delta} R_{\delta \mu}{}^{\alpha \nu} R_{\nu \rho}{}^{\beta \sigma} R_{\sigma \gamma}{}^{\mu \rho}\right] \ , 
\end{aligned}
\end{equation}
for which we wrote  the order $\alpha'{}^3$ contribution in terms of the following function 
\be
J(c) = \left(t_8 t_8 R^4 + \frac{c}{8} e_{10} e_{10}R^4\right) + \text{Ricci terms} \ , \label{JcTypeII}
\ee
where $t_8$ and $e_{10}$ follow the same conventions as in \cite{Codina:2020kvj}. The point of introducing such a function is the following. The term containing $t_8$ can be calculated from four-point scattering amplitudes, whereas the Gauss-Bonnet term with $e_{10}$ starts at fifths order in a field expansion. The cosmological reduction of $J(c)$ in (\ref{JcTypeII}) gives
\be
J(c) = \frac 1 4 (9 - 45 c) (L^8) + \frac 1 {16} (51 + 45 c) (L^4)^2 - 6 (1 - c) (L^3)(L^5) + \L_{\rm II}^{(3)} \ ,
\ee
where $\L_{\rm II}^{(3)}$ contains removable terms that depend on $\dot{L},\dot{\Phi}$ or $(L^2)$, and terms that contribute total derivatives in the action. We then see that the requirement of $O(9,9)$ symmetry fixes the coefficient
 $c$ to its expected value $c = 1$, as it forbids the presence of the interaction $(L^3)(L^5)$. In the subsequent  subsections, dedicated to other theories, this phenomenon will reappear:  
 the couplings contributing to the lowest order scattering amplitudes plus the requirement of duality invariance predicts the coefficient of the Gauss-Bonnet-type  term.

The cosmological reduction of (\ref{LS}) in the form (\ref{action1d}) is then given by
\begin{equation}
\begin{aligned}
L_{\rm II}^{(1)} &= L_{\rm II}^{(2)} = 0 \ ,\\
L_{\rm II}^{(3)} &= \frac{\zeta(3)}{2^{11}}\left[ 
-3(L^8) + 2 (L^4)^2\right] + \L_{\rm II}^{(3)} \ .
\end{aligned}
\end{equation}
The corrections to the EOM  are zero up to order $\alpha'{}^2$, so we have $E_{\Psi}^{(1)} = E_{\Psi}^{(2)} = 0$, and the leading order contribution (\ref{EOM0})  can be used to remove $\L_{\rm II}^{(3)}$ entirely, as explained in Section \ref{sec_general}. The final action can then be written in terms of the generalized metric by using (\ref{SandL}) to arrive at \cite{Codina:2020kvj}
\begin{equation}
S_{\rm II} = \int dt \, e^{-\Phi} \left\{ -\dot \Phi^2  -\frac{1}{8}\tr{\dot {\cal S}^2 } +   \alpha'{}^{3} \frac{\zeta(3)}{2^{12}}\left[ 
-3\, \tr{\dot {\cal S}^8} +    
\tr{\dot {\cal S}^4}^2 \right]\right\} \ .
\end{equation}

\subsection{Bosonic strings}

For the bosonic string, the $26$-dimensional action for the purely metric sector up to and including order $\alpha'{}^2$ was obtained in \cite{Metsaev:1986yb}, based on the  string 3- and 4-point amplitude calculations. It was later extended to include the dilaton contribution in \cite{Jack:1988rq} from the 3-loop metric beta function and a consistency condition proposed in \cite{Curci:1986hiandTseytlin:1986ws}. Finally, the $\alpha'{}^3$ action for the purely metric sector was determined  in \cite{Jack:1989vp} from the 4-loop beta function. 

In \cite{Jack:1989vp} two different schemes were used. Even though the order $\alpha'{}^3$ action was obtained only for the metric sector, both schemes contain terms involving the dilaton, Ricci tensors and Ricci scalars at intermediate orders. Therefore, we found it useful to present the result in an alternative scheme in which those contributions can be redefined away at the expense of changing the $\alpha'{}^3$ couplings. In the main scheme used in \cite{Jack:1989vp} the Lagrangians are  given by
\begin{align}
\mathcal{L'}_{B}^{(1)} &= \frac{1}{4} R_{\mu \nu \rho \sigma} R^{\mu \nu \rho \sigma} \ , \label{LB'1}\\
\mathcal{L'}_{B}^{(2)} &= \frac{1}{16} R_{\mu \nu}{}^{\alpha \beta} R_{\alpha \beta}{}^{\rho \sigma} R_{\rho \sigma}{}^{\mu \nu} - \frac{1}{12} R_{\mu \nu}{}^{\alpha \beta} R_{\alpha \rho}{}^{\mu \sigma} R_{\beta}{}^{\rho \nu}{}_{\sigma} \nn \\
& \quad + \frac{3}{4} \left( R_{\mu \nu} + 2 \nabla_\mu \nabla_\nu \phi\right) R^{\mu \rho \sigma \lambda} R^{\nu}{}_{\rho \sigma \lambda} \nn \\
&\quad + \frac{1}{2}\left( \nabla_\mu \phi \nabla^\mu \phi - \nabla_\mu \nabla^\mu \phi - \frac{1}{4} R\right) R_{\nu \rho \sigma \lambda} R^{\nu \rho \sigma \lambda} \ ,\label{LB'2}\\
\mathcal{L'}_{B}^{(3)} &= \frac{1}{32} R^{\alpha \beta \mu \nu} R_{\mu \nu}{}^{\gamma \delta} R_{\alpha \gamma}{}^{\rho \sigma} R_{\rho \sigma \beta \delta } + \frac{5}{16} R_{\mu \nu \alpha \beta} R^{\mu \nu \alpha \lambda} R_{\lambda \delta \rho \sigma} R^{\beta \delta \rho \sigma} \nn \\
& \quad - \frac{1}{32} R_{\mu \nu \alpha \beta} R^{\mu \nu \alpha \beta} R_{\rho \sigma \gamma \delta} R^{\rho \sigma \gamma \delta} + \mathcal{L}_{\rm II}^{(3)} \ , \label{LB'3}
\end{align}
where the terms $\mathcal{L}_{\rm II}^{(3)}$  are exactly those of  type II string theory, 
with a coefficient proportional to the transcendental $\zeta(3)$ 
that is the same for  all string theories.

Next we perform $D$-dimensional field redefinitions to clean all Ricci and dilaton terms at second order by inducing new contributions at order $\alpha'{}^3$. To do so one needs the corrected EOM that can be rewritten in the following equivalent form
\begin{align}
R_{\mu \nu} + 2 \nabla_\mu \nabla_\nu \phi &= - \alpha' \frac{1}{2} R_{\mu}{}^{\rho \sigma \lambda} R_{\nu \rho \sigma \lambda} + \dots \nn \ , \\
\nabla_\mu \phi \nabla^\mu \phi - \nabla_\mu \nabla^\mu \phi - \frac{1}{4} R &= \alpha' \frac{1}{16} R_{\mu \nu \rho \sigma} R^{\mu \nu \rho \sigma} + \dots \ , \label{rules}
\end{align}
where $\dots$ stands for higher order contributions as well as for dilaton terms at order $\alpha'$. Note that the effect of non-linear variations due to the second order field-redefinitions will start at order $\alpha'{}^4$ and so can be ignored here.  In our case, by using (\ref{rules}) in (\ref{LB'2}), dilaton terms will appear just at order $\alpha'{}^3$. Since $\mathcal{L'}_{B}^{(3)}$ was obtained just for the metric sector, it would be inconsistent to keep induced dilaton terms. In any case, at this order dilaton contributions cannot shift the coefficients of the duality invariant action, as was explained in \cite{Codina:2020kvj}. By applying (\ref{rules}) in (\ref{LB'2}) we can get rid of the dilaton and Ricci terms at the expense of inducing at order $\alpha'{}^3$ two pure Riemann structures together with dilaton terms that we omit. The resulting action in this new scheme is given by
\begin{align}
\mathcal{L}_B^{(1)} &= \frac{1}{4} R_{\mu \nu \rho \sigma} R^{\mu \nu \rho \sigma} \ , \label{LB11}\\
\mathcal{L}_B^{(2)} &= \frac{1}{16} R_{\mu \nu}{}^{\alpha \beta} R_{\alpha \beta}{}^{\rho \sigma} R_{\rho \sigma}{}^{\mu \nu} - \frac{1}{12} R_{\mu \nu}{}^{\alpha \beta} R_{\alpha \rho}{}^{\mu \sigma} R_{\beta}{}^{\rho \nu}{}_{\sigma} \ ,\label{LB22}\\
\mathcal{L}_B^{(3)} &= \frac{1}{32} R^{\alpha \beta \mu \nu} R_{\mu \nu}{}^{\gamma \delta} R_{\alpha \gamma}{}^{\rho \sigma} R_{\rho \sigma \beta \delta } - \frac{1}{16} R_{\mu \nu \alpha \beta} R^{\mu \nu \alpha \lambda} R_{\lambda \delta \rho \sigma} R^{\beta \delta \rho \sigma} + \mathcal{L}_{\rm II}^{(3)} \ . \label{LB33}
\end{align}

At this stage we compactify the action to obtain
\begin{align}
L_B^{(1)} &= \frac{1}{32} (L^4) + \frac{1}{32} (L^2)^2 + \frac{1}{4} (L^2 \dot{L}) + \frac{1}{4}(\dot{L}^2) \ , \label{LB1}\\
L_B^{(2)} &= \frac{1}{192} (L^6) + \frac{1}{16} (\dot{L}^3) - \frac{1}{768} (L^2)^3 + \frac{3}{32} (L^2 \dot{L}^2) + \frac{1}{16} (L^4 \dot{L}) + \frac{1}{256} (L^2)(L^4) - \frac{1}{64} (L^3)(L \dot{L}) \nonumber \\
& \quad - \frac{1}{64}(L \dot{L})^2 + \frac{1}{64} (L \dot{L} L \dot{L}) \ ,\label{LB2} \\
L_B^{(3)} &= \frac{1}{2^{11}}\left[(L^8) - (L^4)^2\right] - \frac{\zeta(3)}{2^{11}}\left[3 (L^8) - 2 (L^4)^2\right] + \L_B^{(3)} \label{LB3} \ ,
\end{align}
where $\L_B^{(3)}$ contains removable terms that depend on $\dot{L},\dot{\Phi}$ or $(L^2)$. 
The EOM then read order by order
\begin{align}
E_g^{(1)} &= \frac{1}{4}\left[\dot{\Phi}^2 - \ddot{\Phi}\right]\left[L^2 + 2 \dot{L}\right] + \frac{1}{8} \dot{\Phi}\left[L^3 + 2 L \dot{L} - 8 \ddot{L} - 6 \dot{L}L +(L^2)L\right]  \\
&-\frac{1}{8}\left[L^2 \dot{L}+L\dot{L}L+\dot{L}L^2\right] + \frac{1}{2}\left[\ddot{L}L + \dddot{L} - L \ddot{L}\right] -\frac{1}{4}(L\dot{L})L - \frac{1}{8}(L^2) \dot{L} \ , \nonumber\\
E_g^{(2)} &= \E_g^{(2)} \ ,\\
E_n^{(1)} &= -\frac{3}{32} (L^4) - \frac{3}{32} (L^2)^2 - \frac{1}{4}(\dot{L}^2) + \frac{1}{2} (L \ddot{L}) - \frac{1}{4} \dot{\Phi}(L^3) - \frac{1}{2} \dot{\Phi} (L\dot{L})\ ,\\
E_n^{(2)} &= -\frac{5}{192} (L^6) + \E_n^{(2)} \ ,\\
E_\Phi^{(1)} &= -\frac{1}{32} (L^4) - \frac{1}{32} (L^2)^2 - \frac{1}{4}(\dot{L}^2) - \frac{1}{4}(L^2 \dot{L}) \ ,\\
E_\Phi^{(2)} &= -\frac{1}{192} (L^6) + \E_\Phi^{(2)} \, , 
\end{align}
where the $\E^{(2)}$ collectively denote removable terms that depend on $\dot{L}, \dot{\Phi}$ and $(L^2)$. 
Using these EOM we can do field redefinitions to bring the action to its minimal form as explained on general grounds in Section \ref{sec_general}. Since this is the first non-trivial example in which higher order EOM are needed for this purpose,  we provide an intermediate step for clarification. When the EOM are used to implement replacements in the first and second order Lagrangians (\ref{LB1}) and (\ref{LB2}), higher order contributions are induced: 
\begin{equation}
\int dt \, e^{-\Phi} \left[\alpha' L_B^{(1)} + \alpha'{}^2 L_B^{(2)}\right] =  \int dt \, e^{-\Phi} \left[\alpha' \frac{1}{32}(L^4) + \alpha'{}^2\frac{1}{192}(L^6) + \alpha'{}^3\frac{1}{2^{14}} (L^4)^2  + \alpha'{}^3 \widetilde{\L}_B{}^{(3)}\right] \, ,
\end{equation}
where in $\widetilde{\L}_B{}^{(3)}$ we collected all removable terms that depend on $\dot{L}, \dot{\Phi}$ and $(L^2)$.  
Note the appearance of a new  $(L^4)^2$ term at order $\alpha'{}^3$. This term  combines with the terms  already present in (\ref{LB3}) and the one that comes from non-linear variations (\ref{L42shift}) with $\gamma = 1$. At this stage any appearance of $\dot{L}, \dot{\Phi}$ and $(L^2)$ is removed by the lowest order EOM. The resulting action is written in terms of traces of even powers of $L$, so it can be cast in terms of the generalized metric using (\ref{SandL}). The manifestly $O(25,25)$ invariant expression is given by
\begin{equation}\label{SB}
\begin{aligned}
S_B = \int dt \, e^{-\Phi} &\left\{ -\dot \Phi^2  -\frac{1}{8}\tr{\dot {\cal S}^2 } + \alpha'\frac{1}{2^6}\tr{\dot {\cal S}^4} - \alpha'{}^2\frac{1}{3.2^7}\tr{\dot {\cal S}^6 } \right.\\
& \quad \left. + \alpha'{}^{3}\frac{1}{2^{12}}\tr{\dot{\S}^8} + \alpha'{}^{3} \frac{1}{2^{16}}\tr{\dot{\S}^4}^2 + \alpha'{}^3\frac{\zeta(3)}{2^{12}}\left[ 
-3\, \tr{\dot {\cal S}^8} +
\tr{\dot {\cal S}^4}^2\right]  \right\} \ .
\end{aligned}
\end{equation}

We conclude this section with a side remark. Notice that we could have deformed the second order contribution  (\ref{LB22}) as follows
\be
\mathcal{L}_B^{(2)} = \frac 1 {48} I_1 + \frac{c}{24} (I_1 - 2 I_2) \ ,
\ee
with 
\begin{equation} \label{I1I2}
\begin{aligned}
I_1 &= R_{\mu \nu}{}^{\alpha \beta} R_{\alpha \beta}{}^{\rho \sigma} R_{\rho \sigma}{}^{\mu \nu}  \ ,\\
I_2 &= R_{\mu \nu}{}^{\alpha \beta} R_{\alpha \rho}{}^{\mu \sigma} R_{\beta}{}^{\rho \nu}{}_{\sigma}  \ ,
\end{aligned}
\end{equation}
such that (\ref{LB22}) is recovered for $c = 1$. Here $I_1$ is the contribution from three-point scattering amplitudes and $I_1 - 2 I_2$ is the cubic Gauss-Bonnet combination arising at quartic powers in a field expansion\cite{Metsaev:1986yb}. Allowing for this freedom, one  encounters the following cosmological action to order $\alpha'{}^2$
\be
S_{B} = \int dt \, e^{-\Phi} \left[ - {\dot \Phi}^2 + \frac 1 4 (L^2) + \frac {\alpha'} {32} (L^4) + \frac {\alpha'{}^2} {768} \left((1 + 3 c) (L^6) +  (1 - c) (L^3)^2 \right) \right] \ .
\ee
The interaction $(L^3)^2$ is not duality invariant so we must take $c = 1$ in order to cancel it. We recognize  here the same behavior found in the Type II action, namely that lowest order scattering amplitudes plus the requirement of duality invariance fixes the coefficient of the Gauss-Bonnet terms.

\subsection{HSZ theory}

The gravitational action of HSZ theory \cite{HSZ} to order $\alpha'{}^3$ only contains a second order contribution
\begin{equation}
\begin{aligned}
\mathcal{L}_{HSZ}^{(1)} &=  0 \ ,\\
\mathcal{L}_{HSZ}^{(2)} &= - \frac 3 4\, \lambda\, \Omega(\Gamma)^2 - \frac 1 {48} I_1 - \frac c {24} (I_1 - 2 I_2) +\cdots 
 \ , \\
\mathcal{L}_{HSZ}^{(3)} &= 0 \, ,
\end{aligned}
\end{equation}
where the dots represent terms with Ricci tensors, Ricci scalars and dilaton couplings. 
The first term in $\mathcal{L}_{HSZ}^{(2)}$ containing the square of the Chern-Simons three-form, 
 \be
 \Omega_{\mu \nu \rho}(\Gamma) = \Gamma^\delta_{[\underline{\mu} \sigma} \partial_{\underline{\nu}} \Gamma^\sigma_{\underline{\rho}]\delta} + \frac 2 3 \Gamma^\delta_{[\underline{\mu} \sigma} \Gamma^\sigma_{\underline{\nu} \lambda} 
    \Gamma^\lambda_{\underline{\rho}] \delta}\;, 
  \ee  
was predicted in \cite{HZGS}, and here we weight it with a coefficient $\lambda$ that should be fixed to $1$ in order to keep track of the terms that it gives rise to in the reduced action.  Without this contribution, the action would be equal to the quadratic bosonic action (\ref{LB22}) up to an overall minus sign. The interactions $I_1$ and $I_2$ were defined in (\ref{I1I2}).
 The second term was computed in \cite{Naseer} through three-point scattering amplitudes, and the cubic Gauss-Bonnet term  was computed in \cite{Lescano}. We weight this contribution with a coefficient $c$ as before, to confirm later that the cosmological reduction will fix its value to $c = 1$, as expected.
 
 HSZ theory follows an interesting patern. Modulo field redefinitions, terms of order ${\cal O}(\alpha'{}^{n})$ with $n$ odd (even) contain odd (even) powers of the two-form \cite{Lescano}. For this reason, purely  gravitational and dilaton terms only appear in orders with even values of $n$. There are then no quadratic nor quartic Riemann interactions in this theory. 

The reduced action takes the form (\ref{action1d}) with
\begin{equation}
\begin{aligned}
L_{HSZ}^{(1)} &=  0 \ ,\\
L_{HSZ}^{(2)} &= - \frac {1} {768} \left((1 + 3 c) (L^6) +  (1 - c) (L^3)^2 \right) + \L_{HSZ}^{(2)} \ , \\
L_{HSZ}^{(3)} &=  0 \ ,
\end{aligned}
\end{equation}
where $\L_{HSZ}^{(2)}$ contains terms that depend on $\dot{L},\dot{\Phi}$ or $(L^2)$. The interaction $(L^3)^2$ is neither duality invariant nor can be eliminated through field redefinitions, so again duality invariance fixes the coefficient of the Gauss-Bonnet term to its expected value $c = 1$. Both the action and the equations of motion contain only second order deformations. For this reason, the terms $\L_{HSZ}^{(2)}$  appearing at second order can be removed to order $\alpha'{}^3$ by using  the leading order EOM (\ref{EOM0}), as explained in Section \ref{sec_general}. 

Cast in a manifestly duality invariant form, the effective action to order $\alpha'{}^3$ then reads 
\be
S_{HSZ} = \int dt \, e^{-\Phi} \left[ - {\dot \Phi}^2 - \frac 1 8 {\rm Tr}(\dot {\cal S}^2)  + \frac {\alpha'{}^2} {384} {\rm Tr}(\dot {\cal S}^6) \right] \ . \label{CosmologicalHSZ}
\ee
Curiously, the absence of $\lambda$ indicates that the Chern-Simons terms leave no trace in the effective cosmological action to this order, 
a fact that is only partially explained by the $\mathbb{Z}_2$ invariance of the cosmological action, c.f.~the discussion in the introduction.

\subsection{Heterotic strings} 

The 10-dimensional low energy effective action for the Heterotic string up to and including order $\alpha'{}^3$ is given by
\begin{align}
\mathcal{L}_H^{(1)} &= \frac{1}{8} R_{\mu \nu \rho \sigma} R^{\mu \nu \rho \sigma} \ , \label{LH1}\\
\mathcal{L}_H^{(2)} &= - \frac{3}{16} \Omega_{\mu \nu \rho} \Omega^{\mu \nu \rho} \ , \label{LH2}\\
\mathcal{L}_H^{(3)} &= \frac{1}{2^{6}} \left[ 18 \Omega^{\mu \nu \rho} \ \Tr{\omega_{\mu} \partial_{\nu} \Omega_{\rho} + \Omega_{\mu}\partial_{\nu} \omega_{\rho} + 2 \Omega_{\mu} \omega_{\nu} \omega_{\rho}} \right.\nonumber\\
& \quad \quad \quad   + 18 R^{\mu \nu \rho \sigma} \Omega_{\rho \mu}{}^{\lambda} \Omega_{\nu \sigma \lambda} + 18 \nabla_{[\mu} \Omega_{\nu] \rho \sigma} \nabla^{\mu} \Omega^{\nu \rho \sigma} \nonumber\\
& \quad \quad \quad - R_{\mu \alpha \beta \gamma} R^{\nu \alpha \beta \gamma} 
R^{\mu \rho \sigma \lambda} R_{\nu \rho \sigma \lambda} - R_{\mu \nu}\,^{\rho \sigma} R_{\rho \sigma}\,^{\alpha \beta} R_{\alpha \beta}\,^{\gamma \delta} R_{\gamma \delta}\,^{\mu \nu}\nonumber\\
& \quad \quad \quad \left. - 2 R_{\mu \sigma}\,^{\alpha \beta} R_{\nu \rho \alpha \beta} R^{\mu \nu \gamma \delta} R_{\gamma \delta}\,^{\rho \sigma} \right] + \mathcal{L}_{\rm II}^{(3)} \, ,
\end{align}
where $\mathcal{L}_{\rm II}^{(3)}$ is defined in (\ref{LS}). 
Up to order $\alpha'{}^2$, (\ref{LH1}) and (\ref{LH2}) coincide with the action calculated in \cite{Metsaev:1986yb} by 3 and 4-point amplitude methods.  Excluding 
$\mathcal{L}_{\rm II}^{(3)}$, we are using the cubic order $\mathcal{L}_H^{(3)}$ obtained in \cite{Bergshoeff:1989de} by supersymmetry. However, the $\alpha'{}^3$ action, including $\mathcal{L}_{\rm II}^{(3)}$, was first found by 4-point scattering amplitude methods in \cite{Cai:1986sa} and \cite{Gross:1986mw}.

In a cosmological background, the theory reduces to
\begin{align}
L_H^{(1)} &= \frac{1}{64} (L^4) + \frac{1}{64} (L^2)^2 + \frac{1}{8} (L^2 \dot{L}) + \frac{1}{8}(\dot{L}^2) \ ,\\
L_H^{(2)} &= \frac{3}{128} (L^2\dot{L}^2) - \frac{3}{128} (L \dot{L} L \dot{L}) \ ,\\
L_H^{(3)} &= - \frac{1}{2^{12}} (L^4)^2 + \frac{\zeta(3)}{2^{11}}\left[-3 (L^8) +2 (L^4)^2\right] + \L_H^{(3)}\ ,
\end{align}
where again we are using the $\L$ notation to indicate terms that are removable by the leading order EOM. To order $\alpha'{}^2$ the deformations to the EOM  are
\begin{align}
E_g^{(1)} &= \frac{1}{8}\left[\dot{\Phi}^2 - \ddot{\Phi}\right]\left[L^2 + 2 \dot{L}\right] + \frac{1}{16} \dot{\Phi}\left[L^3 + 2 L \dot{L} - 8 \ddot{L} - 6 \dot{L}L +(L^2)L\right] \nonumber\\
&-\frac{1}{16}\left[L^2 \dot{L}+L\dot{L}L+\dot{L}L^2\right] + \frac{1}{4}\left[\ddot{L}L + \dddot{L} - L \ddot{L}\right] -\frac{1}{8}(L\dot{L})L - \frac{1}{16}(L^2) \dot{L} \ ,\\
E_g^{(2)} &= \E_g^{(2)} \ ,\\
E_n^{(1)} &= -\frac{3}{64} (L^4) - \frac{3}{64} (L^2)^2 - \frac{1}{8}(\dot{L}^2) + \frac{1}{4} (L \ddot{L}) - \frac{1}{8} \dot{\Phi}(L^3) - \frac{1}{4} \dot{\Phi} (L\dot{L}) \ ,\\
E_n^{(2)} &= \E_n^{(2)}\ ,\\
E_\Phi^{(1)} &= -\frac{1}{64} (L^4) - \frac{1}{64} (L^2)^2 - \frac{1}{8}(\dot{L}^2) - \frac{1}{8}(L^2 \dot{L}) \ ,\\
E_\Phi^{(2)} &=\E_\Phi^{(2)} \ .
\end{align}
Using these $\alpha'$-corrected EOM we can perform field redefinitions and integration by parts to bring the cosmological action to the minimal form. We must not forget the contribution (\ref{L42shift}) arising from non-linear variations in the field redefinitions, where now $\gamma = \frac 1 2$. The resulting action contains traces of only even powers of $L$ so it can be written in a manifestly $O(9,9)$ form using (\ref{SandL}). The final result is
\begin{equation}
\begin{aligned}
S_H = \int dt \, e^{-\Phi} &\left\{ -\dot \Phi^2  -\frac{1}{8}\tr{\dot {\cal S}^2 } + \alpha'\frac{1}{2^7}\tr{\dot {\cal S}^4} - \alpha'{}^{3} \frac{15}{2^{19}}\tr{\dot{\S}^4}^2 \right.\\
& \left. \hspace{3.2cm}  +  \alpha'{}^3\frac{\zeta(3)}{2^{12}}\left[ 
-3\, \tr{\dot {\cal S}^8} +    
\tr{\dot {\cal S}^4}^2\right]  \right\} \ .
\end{aligned}
\end{equation}

\section{Conclusions}

In this paper we have determined the first four coefficients arising in the $O(d,d)$ invariant $\alpha'$ expansion of
string cosmologies for bosonic, heterotic and type II string theory as well as for HSZ theory, completing the result for type II recently announced in \cite{Codina:2020kvj}.
To this end we took the $\alpha'$ corrections to the low-energy effective actions that can be found  in the literature and performed a
cosmological reduction, i.e., assumed that  all fields depend only on time, and then brought these actions
to a canonical field basis. Not for all string theories are the complete higher-dimensional corrections known for metric, B-field and dilaton including up to
eight derivatives, but upon using the $O(d,d)$ symmetry we were able to determine the complete duality invariant cosmological actions 
to order $\alpha'^3$. 
For instance, the complete couplings for the NS-NS fields
at order $\alpha'^3$ remained unknown until the recent proposal in
\cite{Garousi1}, which subsequently has been tested in \cite{Garousi2} by confirming that the full cosmological reduction
agrees with the result obtained in \cite{Codina:2020kvj} from the purely gravitational couplings.

It would be important to cross-check our results by different methods. For instance, one might compute these coefficients by
demanding vanishing of the higher-loop beta functions of a worldsheet theory that is directly adapted to dimensional reduction,
either using a conventional sigma model or an $O(d,d)$ invariant one \cite{Tseytlin:1990va,Bonezzi:2020ryb,Bonezzi:2021mub}.
Alternatively, one may compute string scattering amplitudes in a set-up with $d$-dimensional translation invariance,
which must  be $O(d,d)$ invariant (although the extraction of the cosmological parameters would be somewhat indirect as there
is no scattering in one dimension).

The general classification in \cite{Hohm:2019jgu} characterizes the `space of string cosmologies' to all orders in $\alpha'$,
but since at each order in $\alpha'$ there remain a finite number of parameters that are not determined by $O(d,d)$
it remains an open question to which points in this theory space actual string theories belong.
By determining all free parameters up to and including $\alpha'^3$ we have further restricted the possible subspaces of this theory space
in which the known string theories must live. The general framework of \cite{Hohm:2019jgu}  has already been employed for some investigations
of string inspired cosmology, see e.g.~\cite{Wang:2019kez,Bernardo:2019bkz,Basile:2021amb,Bernardo:2020nol,Edelstein:2020nhg,Nunez:2020hxx,Jonas:2021xkx,Gasperini:2021mat,Quintin:2021eup}, 
and it would be interesting to see if the results presented here  might be useful
for such scenarios.

\subsection*{Acknowledgements} 

We thank Roberto Bonezzi, Felipe Diaz-Jaramillo,  Arkady Tseytlin, Krzysztof Meissner and  Gabriele Veneziano for comments and discussions.

This work  is supported by the European Research Council (ERC) under the European Union's Horizon 2020 research and innovation program (grant agreement No 771862). 
D.~M.~is supported by CONICET. 
T.~C.~is supported by the Deutsche Forschungsgemeinschaft (DFG, German Research Foundation) - Projektnummer 417533893/GRK2575 ``Rethinking Quantum Field Theory".

\end{document}